\documentclass[preprint,12pt, review, authoryear]{elsarticle}

\usepackage{lineno}
\usepackage{amssymb}
\usepackage{amsmath}
\usepackage{booktabs}
\usepackage{microtype}
\usepackage[hidelinks]{hyperref} 
\usepackage{orcidlink}

\usepackage[margin=1in]{geometry}
\usepackage{adjustbox} 

\journal{Advances in Space Research}
\biboptions{sort&compress}

\begin{document}

\begin{frontmatter}
\title{Empirical Evidence of Planetary Group Configurations Modulating Solar Activity} 
\author{Jeffrey A Hansen\, \orcidlink{0009-0000-2251-123X}} 
\affiliation{text={CPM Investing LLC}}

\author{Shaun David Brocus Fell\, \orcidlink{0000-0002-8059-0359}}

\begin{abstract}
Our prior research found that a $90^{\circ}$ configuration of two planetary groups were temporally associated with significant changes in global electromagnetic standing waves (Schumann resonances) during and after the three $90^{\circ}$ configuration events occurring in late 2017 and early 2018. Specifically noted were reductions in variability moving into an event and level changes immediately after a event. Because global standing-wave data have a short history and no $90^{\circ}$ events have occurred since that period, we examine here whether these geometric configuration events correspond to similar signatures in the long sunspot count record. Using daily sunspot data from January 1935 through December 2024, we conducted empirical studies assessing variance changes, post-event level shifts, and potential intrinsic oscillatory structure. In Study A, a variance-ratio test showed that sunspot variability was systematically lower during the events than in the preceding period, with 21 of 26 events since 1935 exhibiting reduced variance (binomial $p=0.0025$). In Study B, solar activity level declined roughly 21 days after events ended, with 21 of 26 events showing negative changes (binomial $p=0.0025$). In Study C, wavelet and filtering analyses revealed no internal solar oscillations at comparable timescales. These findings provide empirical evidence that the configuration events are associated with shifts in solar activity. The next three configuration events in mid- and late-2026 offer an opportunity to assess these patterns in real time.
\end{abstract}
\end{frontmatter}


\section{Introduction} \label{sec1}

\subsection{Purpose of the Study} \label{sec1.1}

This paper examines whether a specific planetary geometry, periods in which the weighted centers of two overlapping planetary groups form an angular separation near $90^{\circ}$, corresponds to measurable short-term changes of solar activity\footnote{The focus on 90° planetary configurations was not motivated by prior astrophysical theory or by an a priori expectation that right-angle geometry would be significant. The angle emerged empirically in earlier work as the common geometric feature associated with statistically unusual timing behavior, without any prior search for that specific geometry. Only subsequently did we note that orthogonal angular relationships are often associated with distinct physical regimes in other areas of astrophysics and plasma physics. This observation is offered solely as contextual background; no physical mechanism is proposed or assumed.}. We use the daily international sunspot record from January 1935 through December 2024 to test for changes during and after these configuration events. 

Our objective is empirical. We test whether sunspot activity exhibits systematic, event-centered changes across the three adjacent intervals, before, during, and after $90^{\circ}$ planetary group configuration events.  We test whether solar activity becomes more stable during these events compared to the immediately prior period, and whether there are systematic changes in the level of solar activity immediately after the events. Sunspot counts provide a long time series that is largely independent of terrestrial electromagnetic conditions and are therefore suitable for examining conditional statistical structure in solar activity.

We do not attempt to identify or model physical mechanisms. While gravitational, tidal, and electromagnetic interactions are mentioned in the literature, existing estimates of their magnitudes are insufficient to explain the observed signatures, and the statistical analyses presented here do not rely on any proposed mechanism.

\subsection{Motivation and Prior Empirical Observations} \label{sec1.2}

In earlier work~\cite{Hansen2025}, we found evidence of specific patterns in global electromagnetic standing waves (Schumann resonances), coinciding with the three configuration events that occurred in late 2017 and early 2018. Because the available Schumann-resonance record is short and no additional configuration events have occurred since 2018, these observations motivated the present investigation into whether the same geometric configurations correspond to statistically significant changes in solar activity. Here we focus exclusively on the daily sunspot record, which provides a much longer observational baseline and allows us to evaluate whether these configurations are associated with recurring changes in solar variability.

The prior research weighted center for the inner group (G1)—Mercury through Saturn—used weights approximating each planet’s relative contribution as suggested by the standard tidal-force proportionality $M/d^3$. The weighted center of the outer group (G2)—Jupiter through Neptune—used weights determined in earlier research. 

The present study focuses solely on the statistical relationship between these $90^{\circ}$ configuration events and solar activity. The findings of the earlier research provide empirical motivation, but no mechanistic interpretation is used in the analyses that follow. 

\subsection{Conceptual Background: Planetary Geometry and Solar Variability }

Planetary orbital geometries influencing solar processes has been an active area of study for years with publications either confirming or denying such a relationship~\cite{DeJager2005,OKAL1975,Callebaut2012,Cameron2013,Cionco2015,Heredia2019, Charbonneau2022, Muoz2023, Scafetta2023, Nataf2022}. The underlying mechanisms usually discussed include planetary tidal forcing, tide-induced resonant excitations, spin-orbit coupling, tachocline torque amplification, among others. The purpose of this study is to contribute to this discussion by highlighting statistical relationships between planetary geometries and solar activity.
 
\subsection{Scope and Limitations of the Present Study }

The focus of this paper is on empirical tests. We make no effort in proposing physical mechanisms, instead only focusing on whether the data exhibits consistent structure conditioned by planetary geometry. 

\subsection{Research Questions and Hypotheses} \label{sec1.3}

This paper evaluates two empirical hypotheses concerning solar activity during the configuration events.

\textbf{Hypothesis A: Solar activity becomes more stable during the $90^{\circ}$ configuration events.} \newline
We test this by comparing variance ratios of sunspot counts for the during-event intervals relative to the 84-day (12-week) periods immediately before each event. A variance ratio below one is interpreted as evidence of increased stability in the during-event window.

\textbf{Hypothesis B: Solar activity exhibits a systematic decline after $90^{\circ}$ configuration events.} \newline
We test this by measuring the level of sunspot counts over the 84-days (12 week) periods after the event. Levels lower than the averages over the last 10-days of the events across a high proportion of events are interpreted as evidence of a systematic change.  

Together, these hypotheses allow us to assess whether the configuration events are associated with statistically detectable changes in solar activity. 

\subsection{Upcoming Configuration Events and Opportunity for Real-Time Testing}

The timing of this paper is important. NASA ephemerides indicate that the next three configuration events will occur in mid- to late-2026. This provides an opportunity for real-time evaluation of the empirical patterns identified here, including whether solar activity exhibits the same stabilization features observed historically.


\section{Literature Context} \label{sec2}

The relationship between planetary configuration and solar activity is contentious. The prevailing view in solar physics holds that planetary gravitational and electromagnetic forces are too weak to influence solar dynamics in a measurable way~\cite{Callebaut2012, Charbonneau2010}. However, a minor line of research has reported correlations between planetary alignments and solar-cycle parameters, including modulation of sunspot activity~\cite{Hung2007,Alabdulgader2018,atmos12030306,Mayrovitz2023}. The present study aligns with the latter body of work, without assuming that any specific physical mechanism is responsible for observed patterns.


\section{Method} \label{sec3}

This section summarizes the key methodological elements used across Studies A through C. Full methodological details are provided within each study. 

Solar activity was measured using the SILSO sunspot series, normalized to a 0.0 to 1.0 scale. Planetary coordinates were obtained from NASA JPL Horizons using consistent settings for reference frame and time scale. The $90^{\circ}$ configuration was defined as the period during which the weighted centers of overlapping inner and outer planetary groups were separated by an angle between $86^{\circ}$ and $94^{\circ}$, with the Sun at the vertex. Variance ratios were calculated for two windows: the 84 days prior to the configuration and the event period itself. The post-event level changes were calculated between the average of the final 10 days of the event and levels over the 84 days following the event. 

Bootstrapping procedures were used to build reference distributions for expected variance ratios, excluding all known configuration periods to avoid contamination. The post-event level changes were compared to levels across the full 89-year sunspot series. Wavelet analysis was performed using a Morlet mother wavelet with $\omega_0 = 6$ to evaluate internal solar oscillations across scales. 

\subsection{Data Sources} \label{sec3.1}

\subsubsection{Sunspot Data} \label{sec3.1.1}
Solar activity was measured using the daily International Sunspot Number (Version 2) published by SILSO~\cite{SILSO}. The raw daily series was normalized to a $[0,1]$-range, using the formula
\begin{equation}
    \bar{d_i} = \frac{d_i - m}{M-m} \; ,
\end{equation}
where $d_i$ is an element of the raw daily sunspot count data set and $m$ and $M$ are the minimum and maximum daily sunspot count across the entire data set, respectively. No smoothing was applied except where explicitly described (Study B). The normalized series provides a consistent basis for comparing activity levels across the 1935-2024 interval.

\subsubsection{Planetary Ephemerides} \label{sec3.1.2}
The underlying planetary coordinates were obtained from the NASA Jet Propulsion Laboratory’s Horizons system, which provides ephemerides calculated from high-precision numerical integrations of solar system dynamics~\cite{JPL_Horizons}. For this study, the daily three dimensional coordinates of the eight major planets were retrieved using consistent settings for reference frame, time scale, and coordinate system. Across all studies, we used the ecliptic coordinate system referenced to the J2000 epoch. These coordinates were used to compute the angular separation between the weighted centers of the inner and outer planetary groups.

\subsection{Definition of the $90^{\circ}$ Configuration Events} \label{sec3.2}

\subsubsection{Planetary Groups and Their Calculated Centers} \label{sec3.2.1}
Two overlapping planetary groups were defined on the basis of their relevance to earlier empirical work: 
\begin{itemize}
    \item \textbf{Inner Group (G1)}: Mercury, Venus, Earth, Jupiter, Saturn
    \item \textbf{Outer Group (G2)}: Jupiter, Saturn, Uranus, Neptune
\end{itemize}

For each group, a weighted mean xyz coordinate was calculated using the daily heliocentric ecliptic datasets obtained from NASA JPL Horizons. These coordinates were then used to compute the weighted centers of the two overlapping groups of planets. 

The weighting scheme used to calculate the center of the inner-planet group (G1) are the rounded values derived from the standard formula for tidal forces ($M/d^3$) \footnote{Mars is excluded from the G1 group because its tidal weighting contributes at the 0.4\% level compared to the other G1 planets.}.  Others have speculated that tidal forces are associated with the level of sunspot counts and solar flares~\cite{Hung2007,Shirley2023}.   

For the outer-planet group (G2), we adopt a weighting scheme defined in earlier work and treat it here as a fixed, ex ante component of the configuration definition. The weights are static over time and are not estimated from, optimized against, or adjusted using sunspot data. In the present study, they serve only to define a reproducible geometric metric and the associated timing of $90^{\circ}$ configuration events. 

Because the weighting G2 scheme was specified prior to the analyses reported here, it does not introduce additional degrees of freedom into the statistical tests. No physical interpretation of the G2 weights is assumed in this study. The statistical analyses evaluate whether sunspot variability differs across configuration-defined periods, independent of the rationale originally motivating the weighting scheme.

Alternative weighting schemes are possible; however, the objective here is not to infer a unique physical weighting but to test whether a pre-defined geometric event set coincides with statistically detectable changes in solar activity.

The final weights are presented in Table~\ref{tab:planetary_weights}. 

\begin{table}[h!]
\centering
\begin{tabular}{lcc}
\toprule
\textbf{Planet} & \textbf{Weight in Inner Group (G1)} & \textbf{Weight in Outer Group (G2)} \\ 
\midrule
Mercury & 14\% & --- \\
Venus   & 34\% & --- \\
Earth   & 15\% & --- \\
Mars    & ---  & --- \\
Jupiter & 35\% & 35\% \\
Saturn  & 2\%  & 40\% \\
Uranus  & ---  & 12.5\% \\
Neptune & ---  & 12.5\% \\
\textbf{Total} & \textbf{100\%} & \textbf{100\%} \\
\bottomrule
\end{tabular}
\caption{Weights used to calculate the center of mass for the inner (Mercury, Venus, Earth, Jupiter, Saturn)  
and outer (Jupiter, Saturn, Uranus, Neptune) planetary groups.  
The inner group center was calculated using standard tidal force proportionality ($M/d^{3}$).  
The outer group weights were determined in prior work, not to optimize predictive performance. 
Solar ephemerides sourced from NASA JPL Horizons.  
}
\label{tab:planetary_weights}
\end{table}

While we believe that the weighting scheme described in Table 1 will create a reasonably accurate forecast for the three 2026 $90^{\circ}$ configuration events, the scheme is unlikely to reflect a highly accurate representation of any true relative influences of orbital geometry and the true orthogonal relationship between the two calculated centers. As examples, the true relative impacts of Uranus and Neptune are unlikely to be equal. The current scheme includes Jupiter in both groups, and an alternative scheme that carves Jupiter from each group and then describes a three-point dynamic involving the inner planets, Jupiter, and the outer planets may be more explanatory. While we note deficiencies in the current scheme, we believe that its inaccuracies would weaken the reported statistical results rather than strengthen them. 

We begin in January 1935 because, in earlier decades, the weighted centers of the two groups approach the Sun closely enough that the inferred angle between the centers becomes poorly conditioned, meaning small differences in ephemerides or weighting assumptions can produce disproportionately large changes in the measured angle. Restricting the analysis to 1935 onward ensures that the $90^{\circ}$ configuration timing is defined by a numerically stable geometric relationship.

\subsubsection{Defining the $90^{\circ}$ Configuration Event} \label{sec3.2.3}

The specific planetary geometry we focus on is the angular separation between the centers of the two  planetary groups. The $86^{\circ}$ to $94^{\circ}$ range allows analysis of a few weeks close to the $90^{\circ}$ mark.  The Sun is located at the vertex of the angle, as enforced by the ecliptic sun-centered coordinate system. The four-degree span before and after the $90^{\circ}$ mark was selected arbitrarily.  

The centers of the G1 and G2 group are computed from the specified weights in Table~\ref{tab:planetary_weights} using the formula
\begin{equation}
    \vec{r}_{1/2}(t) = \sum_{i\in G1/G2} w_i \vec{r}_{\text{planet}_i}(t) \; .
\end{equation}
 Here, $\vec{r}_1$ refers to the three dimensional vector describing the G1 group center with weights $w_{i \in G1}$ and similarly, $\vec{r}_2$ describes the center of the G2 group with weights $w_{i \in G2}$. The time evolution comes from the time evolving ephemerides of each individual planet, $\vec{r}_{\text{planet}_i}(t)$.

We hence define the \textbf{$90^{\circ}$} planetary configuration as any contiguous period during which the absolute angular separation

\begin{equation}
    \theta(t) = \arccos\left( \frac{ \vec{r}_1(t) \cdot \vec{r}_2(t) }{ ||\vec{r}_1(t)|| ||\vec{r}_2(t)|| } \right)
\end{equation}
lies within the range $[86^{\circ}, 94^{\circ}]$. This $8^{\circ}$ window centers the event on the $90^{\circ}$ geometry while allowing for daily ephemeris granularity and a smooth change in angular separation. Each qualifying interval is treated as one event. Henceforth, these $90^{\circ}$ configurations will be called 'events.' Twenty-six events occurred over the 1935 through 2024 period.

For each event, solar activity was partitioned into three windows:
\begin{enumerate}
    \item \textbf{Pre-event}: 84 days before the event
    \item \textbf{During-event}: all days satisfying the $[86^{\circ}, 94^{\circ}]$ window constraint
    \item \textbf{Post-event}: 84 days after the event 
\end{enumerate}

\subsection{Solar Activity Metrics Used in the Analyses} \label{sec3.3}

\subsubsection{Variance Windows (Study A)} \label{sec3.3.1}

Variance was computed as a population variance
\begin{equation}
    \sigma^2_{pre/during} = \frac{1}{N} \sum_{i=1}^{N} \left | d_i - \langle d \rangle \right|^2 \;,
\end{equation}
where $N$ is the number of data points in the pre or during windows, $d_i$ is an indexed data point of the solar activity within the given window, and $\langle d \rangle$ is the mean of the data within the window.

Variance ratios were then constructed using the formula:
\begin{equation} \label{eq:varratios}
    r_{\text{pre}} = \frac{ \sigma^2_{\text{during}} }{ \sigma^2_{\text{pre}} } \;.
\end{equation}

\subsubsection{Symmetric Ten-Day Mean (Study B)}
To reduce day-to-day noise, a centered 10-day window was used for the statistical computations:
\begin{equation}
    \langle d_i \rangle_{10} = \frac{1}{11} \sum_{j=-5}^{5} d_{i+j} \;.
\end{equation}

This smoothed series was used to evaluate level changes at offsets of $0$ to $+70$ days after each event, with $+21$ days producing the strongest consistent pattern across events, as shown in Study B.

\subsection{Bootstrapping and Null Distributions}
To quantify the likelihood of variance-ratio results arising by chance, we constructed two null distributions: 
\begin{itemize}
    \item \textbf{Distribution of individual variance ratios}: The statistical distribution of Eq.~\ref{eq:varratios} is generated by randomly sampling a fixed 42-day window (the median duration of all observed events) from the dataset (excluding the known event windows) and letting the random sample represent a 'during' window. Then the ratios for this random sample are computed, using the 84-day lookback time. This procedure is repeated 50,000 times to generate the null distribution of the ratio statistics.
    \item \textbf{Distribution of sample means}:  26 random variance ratios are computed just like in the previous distribution. The mean of these 26 ratios is computed, representing one sample of the mean. This process is repeated 50,000 times to determine the statistical distribution of the mean statistic. 
\end{itemize}

Percentiles of observed ratios relative to the null distributions were interpreted as empirical one-sided p-scores under the alternative hypothesis of reduced variance during events.

\subsection{Overlap Filtering}
Events were excluded from certain analyses if pre- or post-windows intersected the during-window of adjacent events. The filtering rules thus used are
\begin{itemize}
    \item For $r_{pre}$: Exclude event if another event ends within its 84-day pre-window.
    \item For $r_{post}$: Exclude event if another event begins within its 84-day post-window 
\end{itemize}
This yields a filtered sample of 21 non-overlapping events for studies A and B.

\subsection{Wavelet Decomposition (Study C)}

To test whether internal solar processes operate on event-length timescales, we applied a continuous wavelet transform using a Morlet mother wavelet with $\omega_0 = 6$. 
The wavelet power spectrum identifies dominant oscillations across time and scale. 

The key objective was to test for power enhancements near the approximate 40–100-day band corresponding to the event durations. 
No such enhancements were observed. 

High-pass filtering (cutoffs from 60 days to 14 years) was used to confirm that the detrended solar-activity distribution remains consistent across filter horizons. 

\subsection{Statistical Test}

The following tests were used:
\begin{itemize}
    \item \textbf{One-sided Binomial Test}: Used to evaluate directional consistency.
    \item \textbf{Percentile-derived p-scores}: Determine significance of a random drawing from the probability distributions.
\end{itemize}
All tests were pre-specified based on the study hypotheses.


\section{Results} \label{sec4}

\subsection{Visualization} \label{sec4.1}

Visual evidence of the correlation between three configuration events and our solar activity metric from June 5th, 2015 to December 31, 2019 is provided in Fig.~\ref{fig:zoomedactivity_select}. Fig.~\ref{fig:zoomedactivity} below shows the 89-year history of the dataset.

\begin{figure}[!ht]
    \centering
    \includegraphics[width=\textwidth]{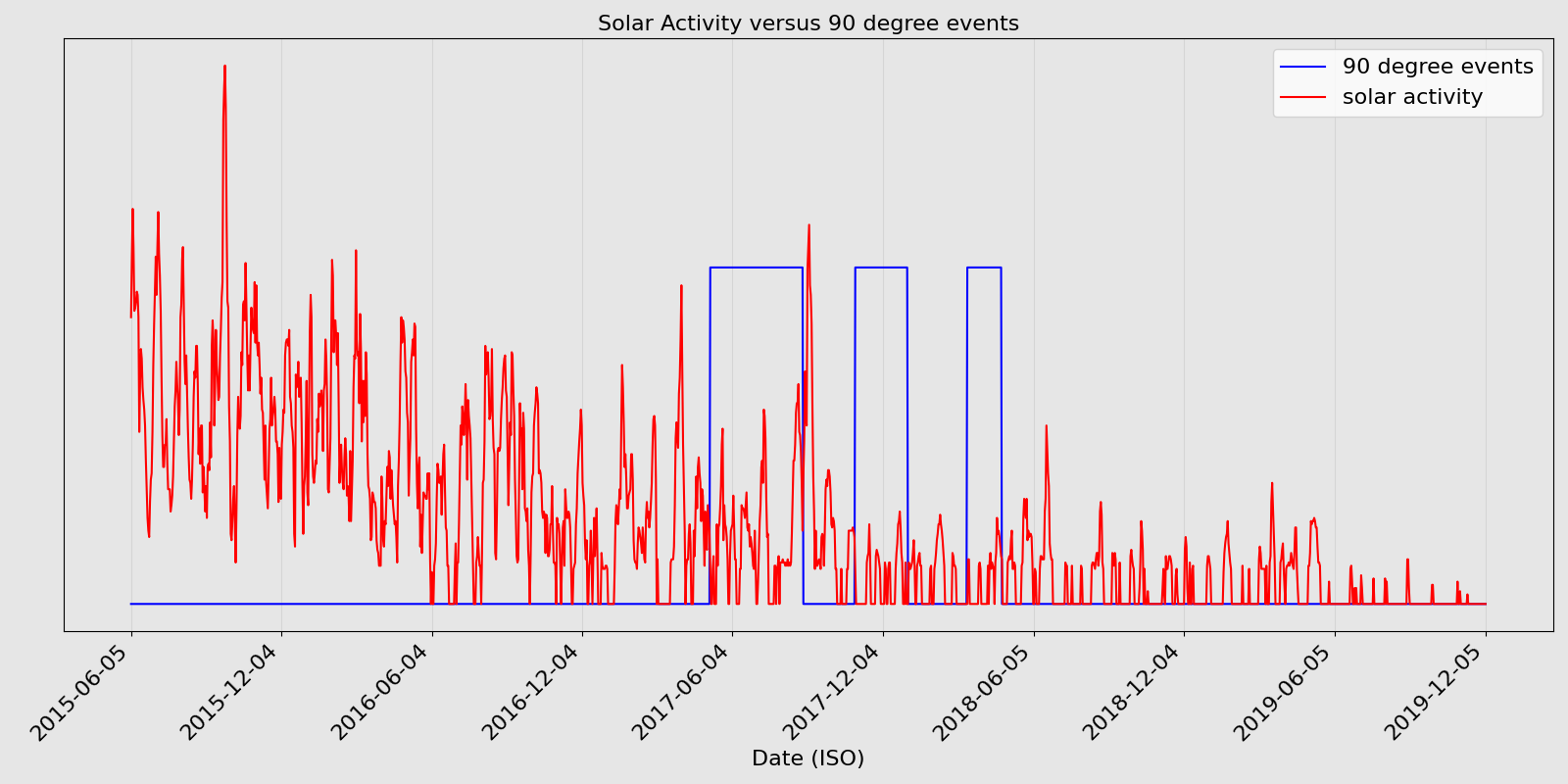}
    \caption{Co-occurrence of solar, and planetary phenomena (2015-2019). The orange line shows the solar activity metric (the normalized daily total sunspot number). Blue bars indicate periods where the angular separation between the inner and outer planetary orbital centers was between $86^{\circ}$ and $94^{\circ}$, as calculated from NASA JPL Horizons ephemeris data. Data sources: Sunspot Number (SILSO data/image, Royal Observatory of Belgium); Planetary Ephemerides (NASA JPL Horizons).}
    \label{fig:zoomedactivity_select}
\end{figure}

\begin{figure}[h!]
    \centering
    \includegraphics[width=\textwidth]{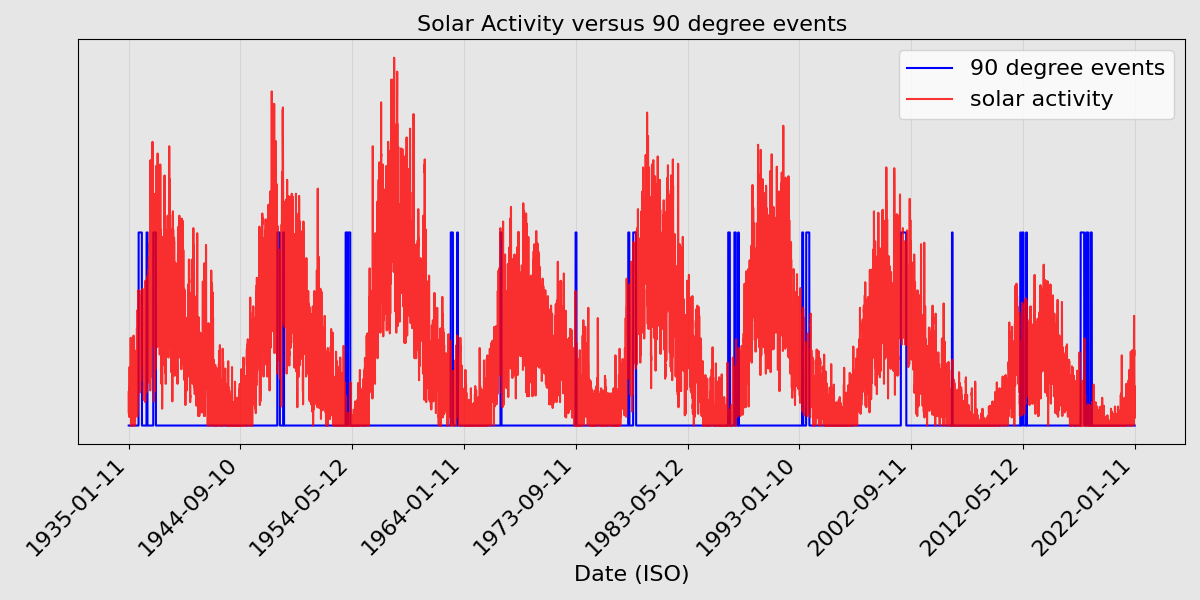}
    \caption{Long-term context of $90^{\circ}$ configuration clusters and solar activity 1935-2025. Vertical blue bands indicate periods where the angular separation between the inner and outer planetary orbital centers was between $86^{\circ}$ and $94^{\circ}$, as calculated from NASA JPL Horizons ephemeris data. Data sources: Sunspot Number (SILSO data/image, Royal Observatory of Belgium); Planetary Ephemerides (NASA JPL Horizons). }
    \label{fig:zoomedactivity}
\end{figure}

There are 26 events over the 89-year period considered in this study. Many of these events are clustered, typically within a 12-month period. There are 13 clusters over the period. The solar activity series exhibits a well-documented ~11-year cycle driven by the Sun’s internal dynamo. The $90^{\circ}$ configurations are not phase-locked to this cycle; they occur at its peaks, troughs, and mid-phases, which suggests their influence is superimposed on the primary 11-year solar rhythm.  

Using the normalized sunspot series as a proxy for the 11-year solar cycle, we tested whether the 26 configuration events tend to occur near solar maxima or minima. For each event, we took the midpoint week and computed its solar activity percentile within the full historical distribution of normalized sunspot values. The resulting percentiles span the full range of the cycle (mean $\sim 0.45$, median $\sim 0.48$), with 13 of 26 events falling above the global median and 5 of 26 in the top quartile. Exact binomial tests under the null of no preference (p = 0.5 for above-median, p = 0.25 for top-quartile) yield p-values of 1.00 and 0.82, respectively, indicating no statistically detectable bias toward any particular phase of the 11-year solar cycle in the timing of these events.

\subsection{Study A --- Variability Before and During Events}

This study involves a test of the variance of solar activity before and during events. We also describe the distribution of the variance ratios and an assessment of the impact of evaluation period overlaps on the test results.  
To that end, we compute the following statistic,
\begin{equation}
r_{\text{pre}} = \frac{\sigma^2_{\text{during}}}{\sigma^2_{\text{pre}}} \;.
\end{equation}
The pre-periods are a fixed window size of 84 days (12 weeks). 

If the $90^{\circ}$ configurations have no impact on solar activity, we would expect ratios centered around 1.00 on average across all 26 events. We find that the median variance ratio $r_{\text{pre}}$ was 0.6528. Twenty-one of 26 events (81\%) exhibited lower variance during the $90^{\circ}$ event than in the 84 days preceding the event (one-sided binomial p=$0.0025$).

\begin{table}[h!]
\centering
\setlength{\tabcolsep}{6pt}
\begin{tabular}{p{4.5cm}c}
\toprule
\textbf{Stability of Solar Activity.} & \textbf{Variance Ratio $r_{\text{pre}}$} \\
\midrule
$n$                              & 26            \\
Median ratio                     & 0.6528        \\
Mean ratio                       & 0.7948        \\
Events with lower variance during & 21/26 = 81\% \\
$p$-score                         & 0.0025       \\
\bottomrule
\end{tabular}
\label{tab:solar_stability}
\end{table}

\subsubsection{Distribution and Samples of the Variance Ratios} \label{sec:4.2.1}

\paragraph{Distribution of Individual Observations}\mbox{}\\
To understand the statistical significance of this ANOVA analysis, we compute the probability density function of the variance ratio statistic over the entire available solar activity data using a bootstrapping process to simulate the population of individual ratios. To prevent contamination of the probability distribution due to the known events in the data, we exclude those date ranges from the random sampling. For the statistic, we take the median duration of the 26 observed events, which is 42 days, and a 84-day lookback period for the pre- and during-event variance ratios. For 50,000 samples, we compute the resulting probability density, shown in Fig.~\ref{fig:ratiodist} below. This probability distribution we henceforth denote by $P^{pre}_{ratio}$ for $r_{\text{pre}}$. 

The $P^{pre}_{ratio}$ population mean is 1.14 with a standard deviation $\sigma$ of 1.27. The distribution has a skewness of 6.49 and an excess kurtosis of 76.33, signifying an extremely right-tailed and tail-heavy distribution.

\begin{figure}
    \centering
    \includegraphics[width=\textwidth]{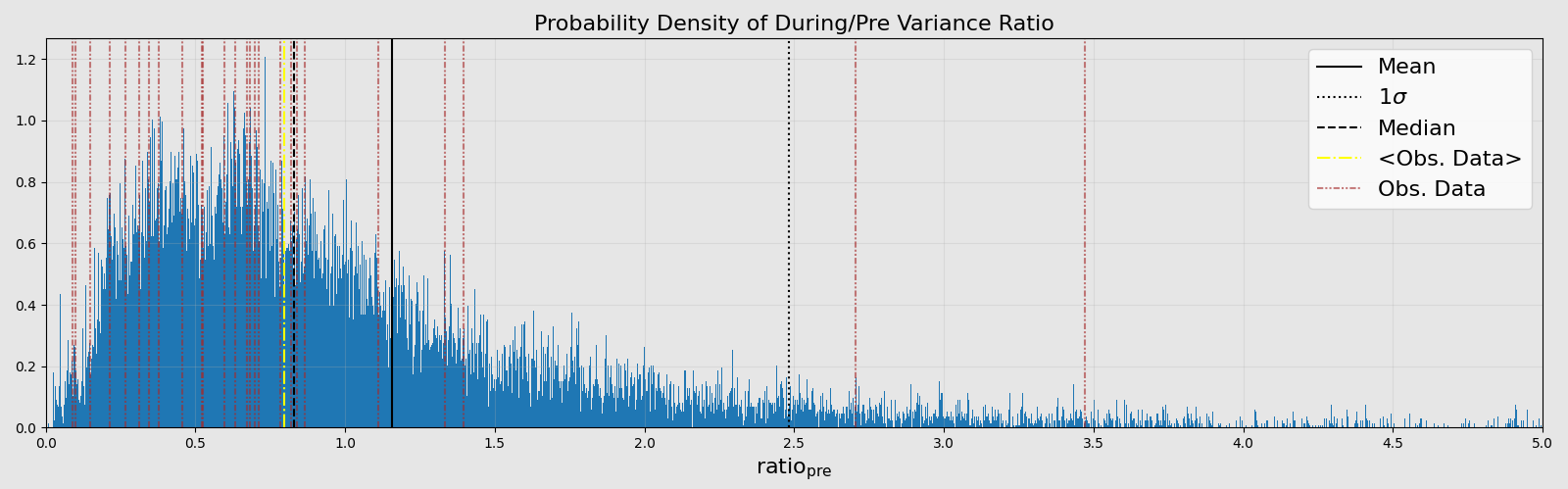}
    \caption{Probability distribution of the pre-variance ratios across the entire dataset. Samples drawn from the dataset explicitly exclude the known $90^{\circ}$ event windows to prevent contamination of the distribution with the known signals. }
    \label{fig:ratiodist}
\end{figure}

The statistical significance of the individual 26 observed ratios can then be determined by computing the percentiles of the observed ratios compared to these distributions. The percentiles of the individual $r_{\text{pre}}$ ratios range from the 1st to the 96th.

The significance is then understood by comparing these quantities using the null and alternative hypothesis. For this study, the alternative hypothesis is that the events correspond to less volatile solar activity data. This implies that the alternative hypothesis corresponds to a during/pre variance ratio \emph{less} than unity. Hence, we should apply a one-sided statistical significance test. This one-sided test corresponds to the probability that a randomly chosen selection of ratios will have a mean that is \emph{less} than the observed ratio. This corresponds precisely to the percentile score.  

For this test, we arrive at individual p-scores for the during/pre variance ratios that range from 0.0116 to 0.9674. 

\paragraph{Distribution of Variance Ratio Means of $N=26$ Samples}\mbox{}\\

We used a similar bootstrapping process to simulate the population of samples of 26 random observations: 

\begin{itemize}
    \item For a single iteration, 26 random choices of the distribution $P^{pre}_{ratio}$ are computed, which forms a vector of length 26 populated by $X \sim P^{pre}_{ratio}$. This vector is denoted by $C^{pre}_{26}$. 
    \item The mean of this vector is computed, representing a single sample of the mean statistic, denoted by $ C^{pre}_{\langle 26 \rangle}$. 
    \item This process is repeated $50,000$ times, building up a vector of length $50,000$ consisting of realizations of the new distribution, $ C^{pre}_{\langle 26 \rangle}  \sim  P^{pre}_{\langle ratio \rangle}$. 
\end{itemize}

The complete probability distribution is then computed from these realizations, displayed in Fig.~\ref{fig:ratiodist_mean} below. This is the distribution of the mean statistic for samples of 26 random variance ratios. We can directly compare the mean $r_{\text{pre}}$ variance ratios of the data against this distribution, also displayed in the figure.  

The skewness of the $P^{pre}_{\langle ratio \rangle}$ distribution is 1.35, with an excess kurtosis of 0.36. 

We find that the mean observed $r_{\text{pre}}$ variance ratio lies within the 3rd percentile. Using the same significance test as above, these percentiles correspond to the p-score itself for our alternative hypothesis. Hence, the mean of the $r_{\text{pre}}$ variance ratio has a p-score of $p=0.03$. This signifies the $r_{\text{pre}}$ variance ratio is moderately significant. These results are consistent with the first test.

\begin{figure}[h!]
    \centering
    \includegraphics[width=\textwidth]{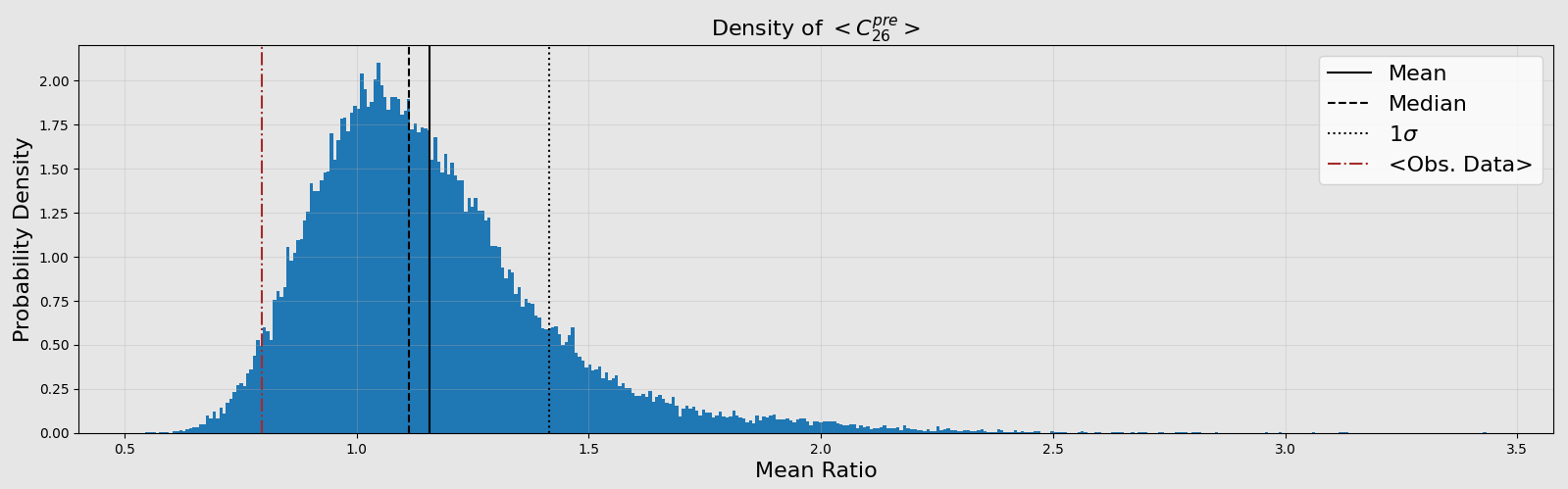}
    \caption{Probability distribution of the mean of 26 randomly sampled pre-variance ratios across the entire dataset. Samples drawn from the dataset explicitly exclude the known $90^{\circ}$ event windows to prevent contamination of the distribution with the known signals. }
    \label{fig:ratiodist_mean}
\end{figure}

\paragraph{Checking the Impact of Overlapping Events}\mbox{}\\

Since some 26 events are temporally clustered, typically within a 12 month period, as shown by the three events in Fig.~\ref{fig:zoomedactivity_select}, the 84-day periods before events may be affected by adjacent events. For completeness, we repeat the above analysis by omitting events in the following way: If an event occurs within 12 weeks after another event, the current event is discarded during the during/pre variance ratio computation.

With these omissions, the total number of events drops from 26 to 21. We repeat the same analysis as above. For the comparison of the observations to the full statistic distribution, the significance does not change. For the comparison of the mean of the ratios to the statistical distribution of the mean statistic, we find slightly less significance. For 21 $r_{\text{pre}}$ variance ratios, we find a p-score for the mean of $p=0.05$. These results are consistent with the earlier tests.

\subsection{Study B --- Patterns in the Post Event Period}

We analyzed the change in solar activity level between the end of the event and a range of lengths of time after the event. To minimize the influence of random daily noise, we average solar activity across a centered 10-day symmetric window. We used the same centered 10-day mean of solar activity for the the end of the forward-looking period. We found that the greatest magnitude of the difference across all events occurs at the +21-day timescale after the end of the event. For this study, we verified that no events overlap with the +21-day period.

A plot of the mean and median change in solar activity between the end of the event and a range of lengths of time after the event is shown in Fig.~\ref{fig:delta_g2pos_vs_lookforward} below. The greatest change across all events occurs roughly at the +21-day timescale with a sharp decline in solar activity. As shown in the following statistical test, the +21-day delta is significantly different from zero making it unlikely to arise from internal processes, suggesting that the alternative hypothesis in this study is more likely.

\begin{figure}[h!]
    \centering
    \includegraphics[width=\textwidth]{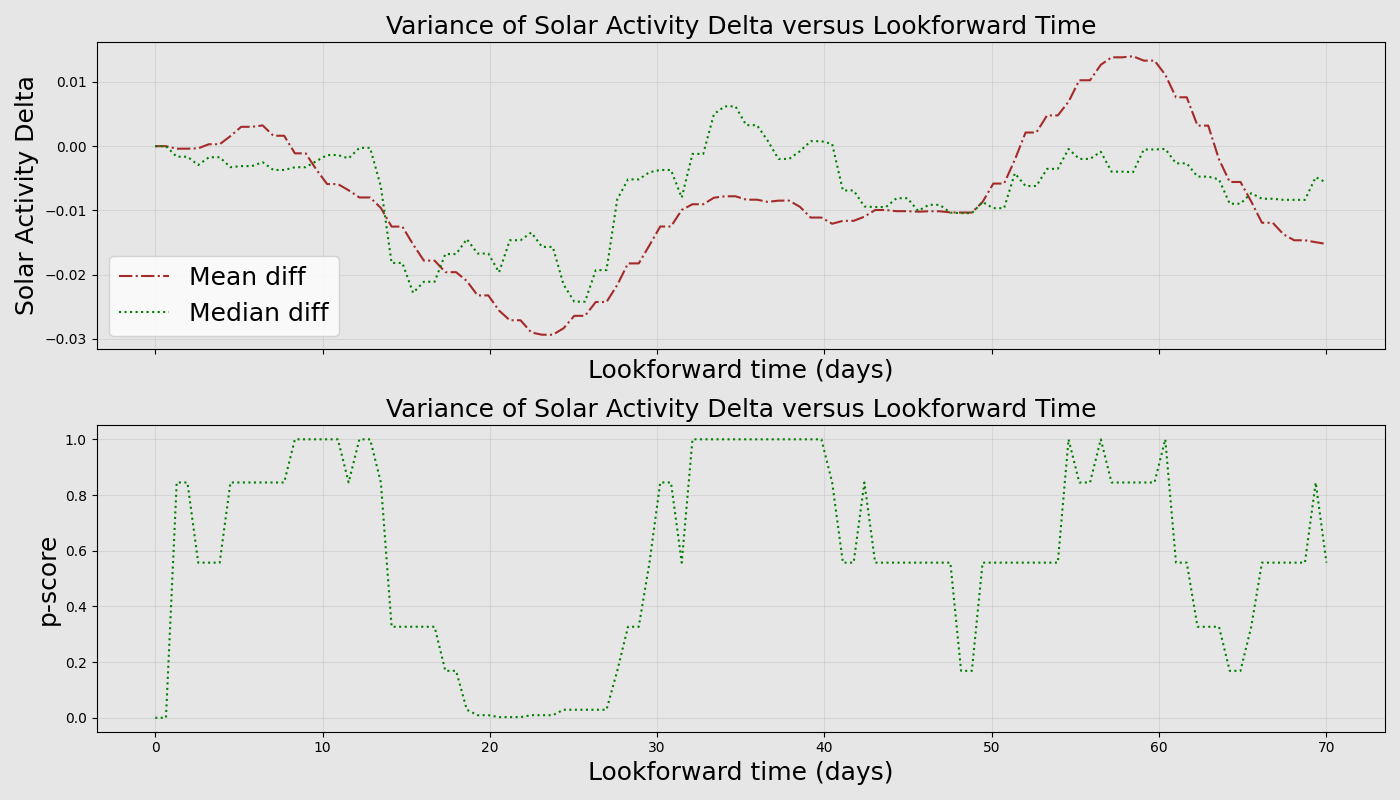}
    \caption{Statistics of the post-event solar activity delta versus lookforward horizon (window size). Bottom plot shows the associated p-value from the number of events with a negative difference (less activity) over the lookforward horizon.}
    \label{fig:delta_g2pos_vs_lookforward}
\end{figure}

\subsection{Study C --- Distinguishing Internal Solar Variability from Configuration Effects}

Study C examines whether the statistical patterns observed in the earlier analyses—especially the reduced variability of solar activity during $90^{\circ}$ configuration events—could simply be artifacts of the Sun’s own internal cycles. The sunspot record contains strong long-term structure, most notably the 11-year Schwabe cycle and a range of slower and faster oscillations. If any of these internal processes produced variability on time scales similar to the 86-day windows used in our variance-ratio tests, then the results of Studies A and B might reflect natural solar dynamics rather than any response to planetary geometry. To evaluate this possibility, we analyze the full distribution of detrended solar activity across multiple filter cutoffs and apply a continuous wavelet transform to identify all dominant time scales in the data. This approach allows us to determine whether known or hidden internal processes could mimic the statistical signals attributed to the $90^{\circ}$ configurations.  

\begin{figure}[h!]
    \centering
    \includegraphics[width=\textwidth]{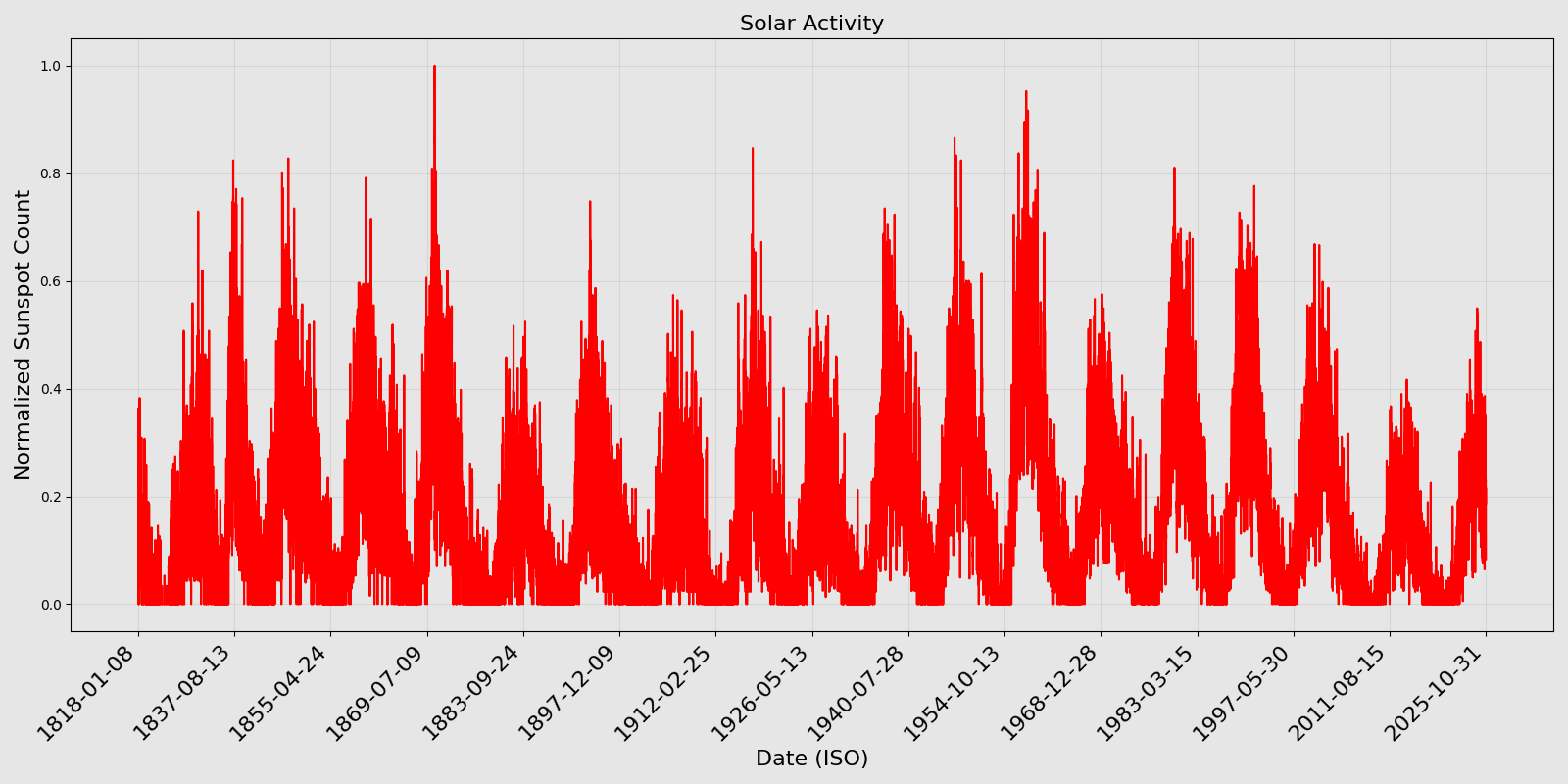}
    \caption{Full solar activity data (normalized to $[0,1]$) from 1818 to 2025}
    \label{fig:full_tranq}
\end{figure}

This study looks at the solar activity probability density to better understand how the various lookback/forward times could affect the statistics by unknowingly partially capturing broader-scale structure in the activity data. The complete set of activity data is shown in Fig.~\ref{fig:full_tranq}. 

\begin{figure}[h!]
    \centering
    \includegraphics[width=\textwidth]{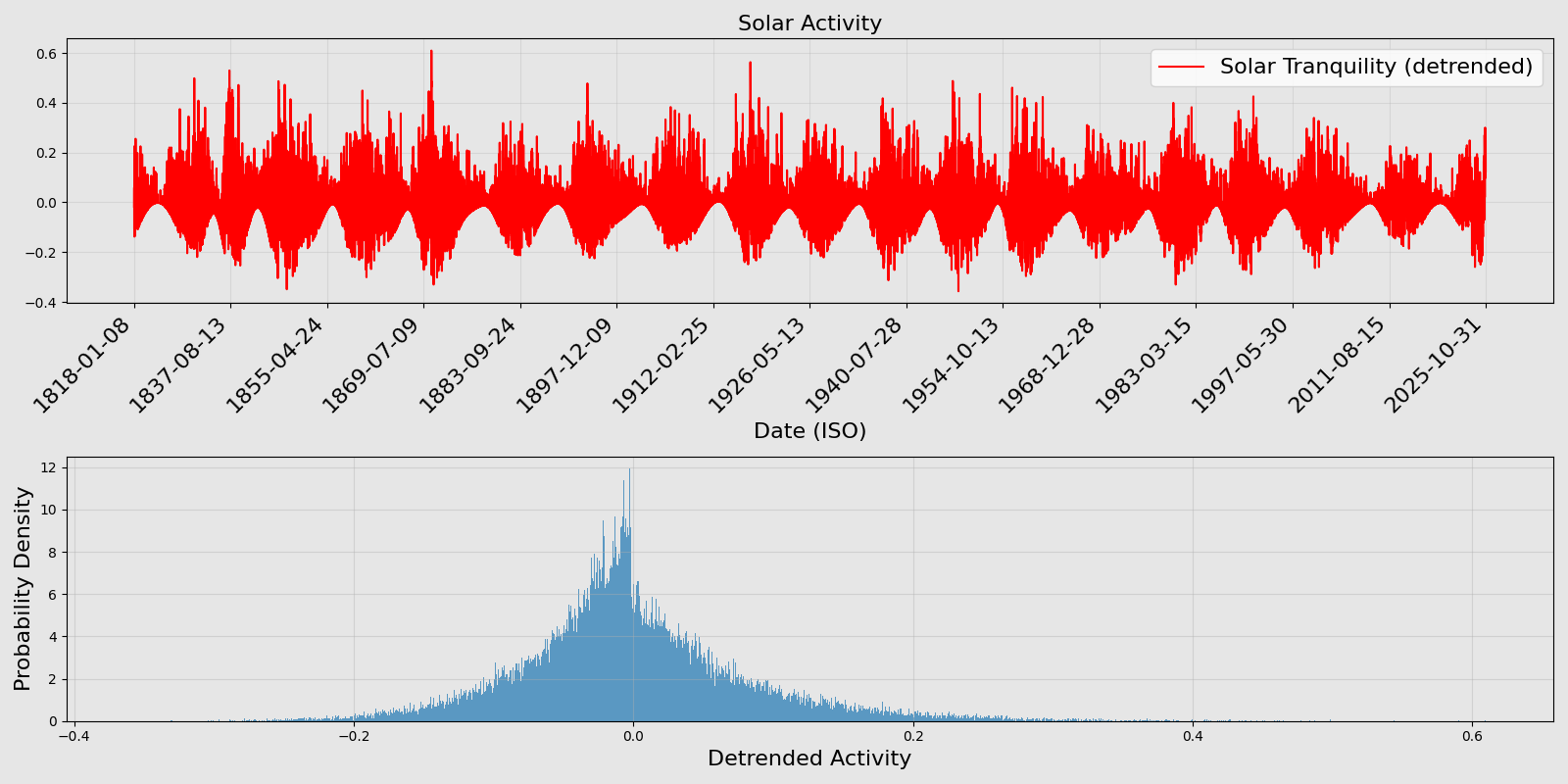}
    \caption{Full solar activity data (normalized to 0-1) from 1818 to 2025. Filtered to remove the dominate 11-year cycle.}
    \label{fig:full_tranq_filter}
\end{figure}

We immediately identify the large-scale 11-year solar cycle. To look at the underlying distribution, we perform a high-pass filter with a cutoff of 9 years, where the cutoff was chosen to sufficiently filter out the dominant 11-year cycle (See Fig.~\ref{fig:full_act_cwt}). The filtered data reveals less dominant periodic signals within solar activity. To that end, we take a Butterworth highpass filter to maximally preserve the passband data. The 9-year cutoff corresponds to a frequency cutoff of $f_c = 0.0003 \frac{1}{\text{day}}$. The detrended activity, along with the resulting distribution, is shown in Fig.~\ref{fig:full_tranq_filter}.  

The distribution mean is 0 with a standard deviation of $0.083$. The distribution has a skewness of $0.731$ and excess kurtosis of $-0.575$. The skewness describes a markedly right-tailed distribution, compared to the mean, implying that the detrended activity has more positive values than negative (heavy right-tailed). Most of the data lies left of the average, favoring less solar activity. The negative excess kurtosis implies that the distribution is platykurtic, meaning the distribution is flatter than a normal distribution and the data is more spread in the center and mid-range rather than extremes. Since the internal solar processes giving rise to the underlying structure are extremely rich and can vary from timescales as short as 154 days to over 11 years, we vary the filter cutoff from 6 months up to 14 years. We find that throughout this range, the distribution metrics remain qualitatively unchanged. The distribution remains left-tailed and platykurtic, independent of the filter cutoff. The Butterworth filter is designed to be maximally flat within the passband, thus any structures present in the data whose characteristic frequencies are larger than the cutoff frequency are largely uneffected by the filtering. Hence, none of these cutoffs substantially effect structures with periods less than 6 months, which includes all the $90^{\circ}$ events. 

Since the distribution remains platykurtic, this implies the variance of the data is not dominated by the extreme values. Since most of the data is centrally located around the median, the variance is determined largely by the most common values. This means the variance ratios we computed in the preceding study are not effected by extreme outliers.

\subsubsection{Statistical Properties of Solar Activity}

In addition to the statistical properties of the solar activity, we aim to characterize the oscillatory content of the data across time and scale. This serves to reinforce the statistical robustness of the previous analyses by demonstrating that the relevant dynamical timescales of the internal solar dynamo and related processes are much longer than the 86-day windows used in the variance-ratio computations. Consequently, these statistics are not contaminated by shorter-timescale solar dynamics. 

We perform a continuous wavelet decomposition using a Morlet mother wavelet with $\omega_0 = 6$. The Morlet wavelet is chosen due to its excellent time-frequency localization, meaning it can identify when a frequency component appears and how long it persists, due to the wavelet scale being almost identical to the Fourier period. This implies oscillatory structure in the solar activity will be much better resolved compared to other wavelets. $\omega_0 = 6$ balances time and frequency resolution~\cite{Torrence1998}. The resulting wavelet power spectrum is displayed in Fig.~\ref{fig:full_act_cwt} below. The 11-year Schwabe cycle is clearly visible as a strong band of power at a period of approximately 11 years, along with subharmonics at longer periods. In the logarithmic plot of the power spectrum, dark vertical streaks appear at periods below one year; these arise from the choice of proxy for solar activity—the sunspot count—which is bounded below at zero. This lower bound corresponds to a normalized solar activity of 0 and produces the characteristic dropout artifacts in the wavelet power. For comparison, the normalized solar activity level is shown in the bottom panel. Additionally, small marks are overlaid on the power spectrum corresponding to the $90^{\circ}$ events since 1935, positioned horizontally at their occurrence times and vertically according to their durations in years. For significance of the power spectrum, the $95\%$ confidence level is also computed and shown in both the power spectrum and its log plot. 

The wavelet spectrum allows us to determine whether additional solar processes could interfere with our statistics. Any such process would manifest as persistent excess power within the period band associated with the duration of the $90^{\circ}$ events. Were this excess present, it would suggest that internal solar variability, rather than external forcing, could explain the observed statistics. However, we find no such enhancement within that band. Since variance relates to the frequency interpretation via Parseval’s theorem, this provides the basis for analyzing the variance ratios in the preceding studies --- frequencies originating from within the solar environment whose characteristic periods are on the order of the duration of the $90^{\circ}$ events would interfere with our alternative hypothesis and the related statistics. Since we do not find excess power in the wavelet power spectrum at these periods, this gives the statistical basis for confirming our alternative hypothesis using the previous variance ratios. In other words, there are no solar processes present at the scales of interest that would interfere with our studies.   

\begin{figure}[h!]
    \centering
    \includegraphics[width=\textwidth]{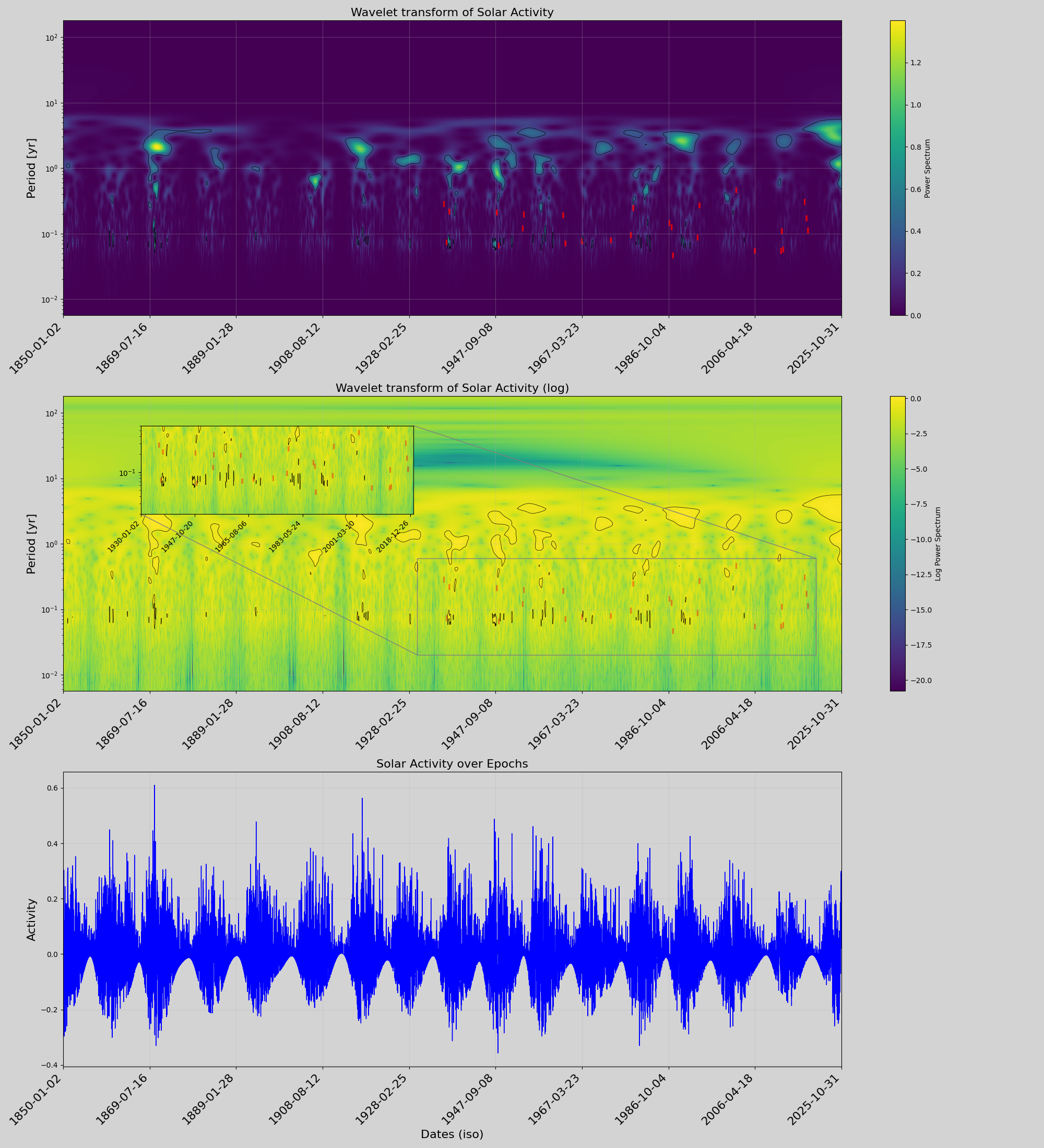}
    \caption{Wavelet analysis of the solar activity time series, filtered to remove the dominate 11-year solar cycle. Top plot shows the complete wavelet power spectrum, while the middle plot shows the log of the power spectrum, to reveal subdominant features. The bottom plot is the (normalized) solar activity time series. Overlayed in the power spectrum plots are the 26 $90^{\circ}$ events. Their vertical position along the 'period' y-axis is determined by the duration of each event.}
    \label{fig:subdom_act_cwt}
\end{figure}

To confirm subdominant modes don’t influence the $90^{\circ}$ event windows, we filter the solar activity data as in the previous study, perform the same wavelet analysis, and analyze the $95\%$ confidence levels. The wavelet analysis of this filtered data is displayed in Fig.~\ref{fig:subdom_act_cwt}. Again, we find no significant power during the $90^{\circ}$ events. 

Our reasoning proceeds as follows: 
\begin{enumerate}
    \item The wavelet power spectrum represents the local contribution of different scales (corresponding to approximate frequencies) in the data as a function of time. 
    \item A local reduction in variability of the underlying signal corresponds to diminished power at the higher (approximate) frequencies; conversely, a local increase in variability produces enhanced power at these frequencies.
    \item The absence of sporadic power enhancements within the period band associated with the $90^{\circ}$ events indicates that no internal solar processes operate on these timescales with sufficient strength to explain the observed reductions in variance. 
    \item Therefore, the observed reduction in the variance of solar activity during the $90^{\circ}$ events is most naturally explained by the forcing from planetary geometry rather than by internal solar dynamics. 
\end{enumerate}

This study confirms the expected results from known solar cycles. The $90^{\circ}$ events predominantly have durations within the 20-100 day range. The Rieger period is one of the shortest solar periods (besides solar rotation rate and granulation) with a timescale from 150-180 days, well outside the duration of the $90^{\circ}$ events~\cite{Hathaway2015}.

\begin{figure}[h!]
    \centering
    \includegraphics[width=\textwidth]{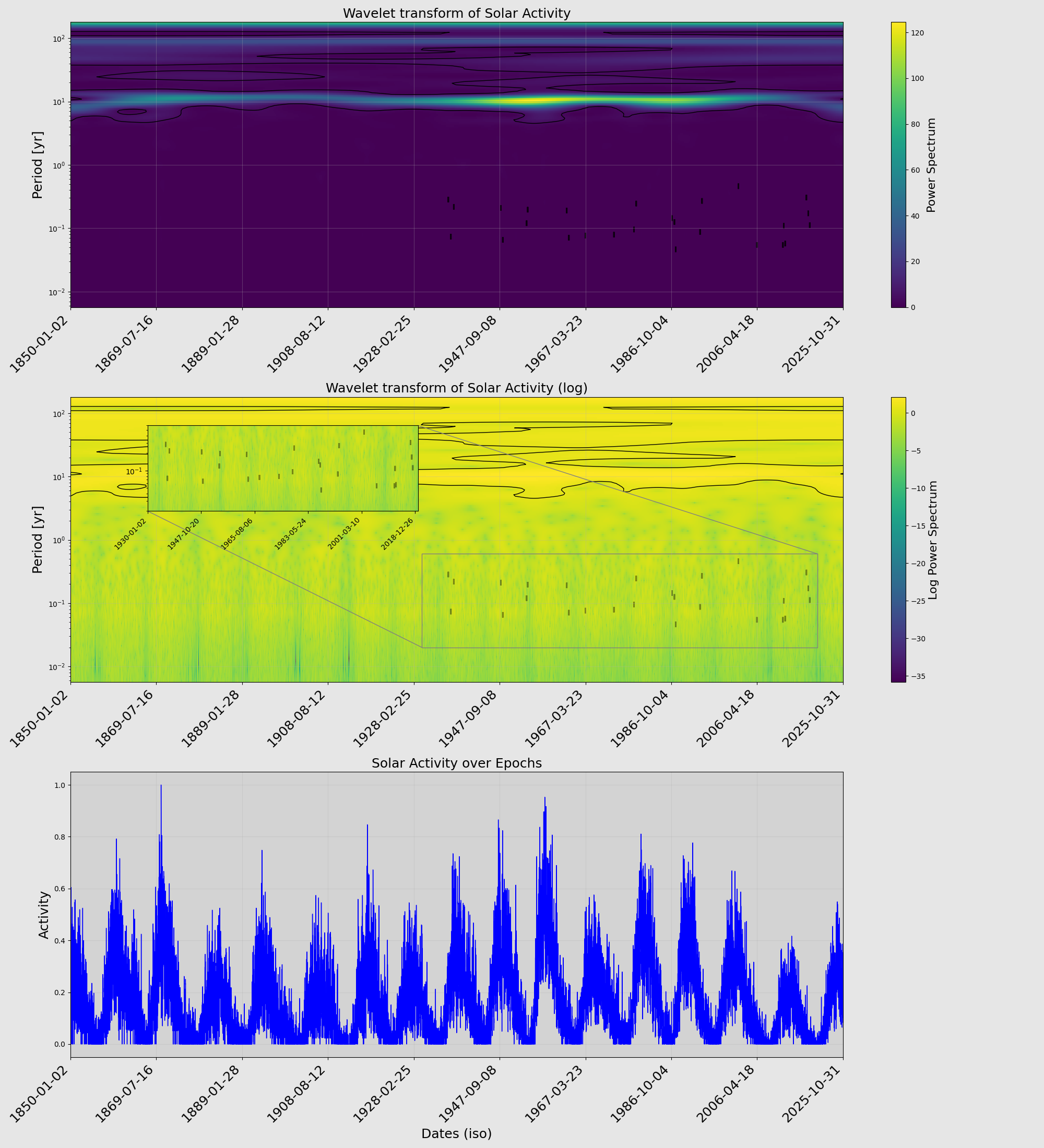}
    \caption{Wavelet analysis of the full solar activity time series. Top plot shows the complete wavelet power spectrum, while the middle plot shows the log of the power spectrum, to reveal subdominant features. The bottom plot is the (normalized) solar activity time series. Overlayed in the power spectrum plots are the 26 $90^{\circ}$ events. Their vertical position along the 'period' y-axis is determined by the duration of each event.}
    \label{fig:full_act_cwt}
\end{figure} 

Taken together, the filtering results and the wavelet power spectrum show that the Sun exhibits no meaningful internal variability on the time scales relevant to our statistical tests. The detrended distribution remains stable regardless of the filtering horizon, and the wavelet analysis reveals no persistent power or structural features in the period band corresponding to the durations of the $90^{\circ}$ events. In particular, there is no evidence of short-term solar dynamics that could generate reductions in variance or post-event shifts similar to those identified in Studies A and B. These findings indicate that the observed patterns are not artifacts of solar-cycle structure or internal dynamical processes. Instead, the empirical signatures of stabilization during $90^{\circ}$ configuration events is consistent with an external conditioning influence rather than internal oscillation.


\section{Discussion} \label{sec5}
The results presented in this paper show that $90^{\circ}$ planetary configuration events correspond to distinct regimes in the historical sunspot record of reduced variability during the event compared to the period immediately prior to the event and to a systematic decline after the event. These findings emerges across multiple independent analyses, time windows, and statistical summaries.

Study A demonstrated that the variance of sunspot activity is systematically lower during the events than in the 84 days (12 weeks) preceding them. 21 of the 26 events exhibit this reduction, producing a binomial p-score of 0.0025. The simulated null distribution places the mean $r_{\text{pre}}$ near the lower tail, indicating that this pattern is unlikely to arise from the normal variability of the sunspot count series. 

Study B identified an a consistent downward shift in solar activity levels approximately 21 days after the end of the $90^{\circ}$ configuration. Using a centered 10-day mean to minimize short-term noise, 21 of the 26 events display negative deltas at this +21-day point, yielding a binomial p-score of 0.0005.   

Study C tested whether internal solar dynamics could generate patterns that resemble those associated with the $90^{\circ}$ configurations. The wavelet analysis showed no meaningful power at periods corresponding to the durations of the events, and filtered distributions were stable across a wide range of cycle-cutoff frequencies. These results indicate that the event-level patterns are not mathematically encoded within the Sun’s internal oscillatory structure. Together with the variance and trend tests, these findings suggest that the observed signatures are not internal artifacts of the solar cycle or shorter-term intrinsic variability. 

Although the number of events is necessarily limited to the historical record, three independent lines of evidence—variance stabilization, post-event level shifts, and the absence of corresponding internal oscillations—lead to a consistent conclusion: the $90^{\circ}$ configuration events mark periods in which the Sun behaves differently. The next cluster of events in 2026 provides an opportunity for direct, real-time testing of these predictions.


\section{Conclusion} \label{sec6}
The analyses in this paper demonstrate that the $90^{\circ}$ planetary configurations correspond to repeatable, statistically significant changes in solar activity. Across nearly nine decades of observations, sunspot variance decreases during these events relative to the prior 84 days and solar activity systematically declines approximately 21 days after the end of each configuration event. Both are unlikely to arise by chance.

Filtering and wavelet analyses confirm that these patterns do not originate from the Sun’s internal oscillatory structure. Instead, the configuration events act as markers for transitions in the dynamics of sunspot activity that are not explainable by cycle phase, internal turbulence, or short-term solar processes. The evidence presented here does not identify the underlying mechanism. The aim of the paper is empirical: to document a set of statistically coherent signatures that occur around well-defined geometric configurations in the solar system. 

These findings challenge the prevailing assumption that planetary geometry cannot influence solar activity. While further theoretical work is required to determine whether a physical mechanism can produce the observed effects, the empirical regularities documented here warrant deeper investigation. The forthcoming sequence of $90^{\circ}$ configurations in 2026 offers an opportunity for real-time evaluation of the predictions implied by this analysis.


\section{Declarations} \label{sec7}
\subsection*{Funding} 
The authors received no financial support for the research, authorship, or publication of this article. 

\subsection*{Conflicts of Interest }
The authors declare no competing financial or non-financial interests. 

\subsection*{Availability of Data and Materials}
All data employed in this study are publicly available. Sunspot records were obtained from the World Data Center SILSO (Royal Observatory of Belgium). Planetary ephemerides were obtained from the NASA JPL Horizons system. Processed datasets used in the analysis are available from~\cite{Fell_SolarPlanetaryCorrelation_2025}. 

\subsection*{Code Availability}
The numerical analyses were carried out using standard statistical routines in Python and Microsoft Excel. Code used to generate the results is available from~\cite{Fell_SolarPlanetaryCorrelation_2025}. 

\subsection*{Ethical Approval}
Not applicable. The study does not involve human participants, animal subjects, or sensitive personal data. 

\subsection*{Consent to Participate / Publish}
Not applicable. 

\subsection*{Author Contributions}
JAH was responsible for the conception of the studies and interpretation of the findings. SDBF performed the data analysis and contributed to the writing and critical revision of the manuscript.

\bibliographystyle{elsarticle-harv}
\bibliography{cas-refs}

\end{document}